\documentclass[11pt]{article}
\pdfoutput=1

\usepackage[english]{babel}
\usepackage{graphicx}
\usepackage{natbib}
\usepackage{amsfonts,amssymb,amsmath,amscd,amsthm,latexsym}
\usepackage{a4wide}
\usepackage[a4paper,citecolor=blue,linkcolor=red,colorlinks=true]{hyperref}
\usepackage{appendix}
\usepackage{fullpage}
\usepackage{setspace} 
\usepackage{geometry}
   
\usepackage{lineno}

\newcommand{\I}{\mathbb{I}}
\newcommand{\E}{\mathbb{E}}
\renewcommand{\P}{\mathbb{P}}

\newcommand{\bbeta}{\boldsymbol{\beta}}

\newcommand{\btheta}{\boldsymbol{\theta}}

\newcommand{\bX}{\mathbf{X}}
\newcommand{\bZ}{\mathbf{Z}}

\newcommand{\bz}{\mathbf{z}}

\newcommand{\bx}{\mathbf{x}}

\newcommand{\bb}{\mathbf{b}}

\usepackage{color}

\newcommand{\Covariance}{\mathbb{C}\mathrm{ov}}

\theoremstyle{plain}

\newtheorem{proposition}{Proposition}[section]
\newtheorem{remarque}{Remarque}[section]

\newcommand{\Paragraph}[1]{{\bf{#1}}\newline} 
\newcommand{\Preuve}{\Paragraph{Proof}}
 \newcommand{\Fin}{$\Box$\\}
\numberwithin{equation}{section}

\title{Bounding rare event probabilities in computer experiments}

\author{
Yves Auffray \\
Dassault Aviation \& \\
D\'epartement de Math\'ematiques, \\ 
Universit\'e Paris-Sud, France
\and	
Pierre Barbillon \\
INRIA Saclay, projet \textsc{select}, \\
D\'epartement de Math\'ematiques, \\ 
Universit\'e Paris-Sud, France
\and
Jean-Michel Marin
\footnote{Corresponding author: place Eug\`ene Bataillon, Case Courrier 051, 34095 Montpellier cedex 5}
\footnote{\textsc{jean-michel.marin@univ-montp2.fr}} \\
Institut de Math\'ematiques et Mod\'elisation de Montpellier \\
Universit\'e Montpellier 2
}

\date{}

\begin{document}

\maketitle

\begin{abstract}

We are interested in bounding probabilities of rare events
in the context of computer experiments. These rare events depend on the output
of a physical model with random input variables. Since the model is only known
through an expensive black box function, standard efficient Monte Carlo methods
designed for rare events cannot be used. We then propose a strategy to deal
with this difficulty based on importance sampling methods. This proposal relies on
Kriging metamodeling and is able to achieve sharp upper confidence bounds on the rare event probabilities.
The variability due to the Kriging metamodeling step is properly taken into account. \\
The proposed methodology is applied to a toy example and compared to more standard Bayesian bounds.
Finally, a challenging real case study is analyzed. It
consists of finding an upper bound of the probability that the trajectory of an
airborne load will collide with the aircraft that has released it. 

\vspace{0.5cm} \noindent \textbf{Keywords:} computer experiments, rare events, Kriging,
importance sampling, Bayesian estimates, risk assessment with fighter aircraft.
\end{abstract}

\newpage

\section{Introduction}
\label{secintroduction}

Rare events are a major concern in the reliability of complex systems \citep{heidelberg:1995,shahabuddin:1995}. 
We focus here on rare events depending on computer experiments.
A computer experiment \citep{welch:etal:1992,koehler:owen:1996} consists of an evaluation of a black box function
which describes a physical model,
\begin{equation}
 \label{computerexp}
y=f(\bx)\,,
\end{equation}
where $y\in \mathbb{R}$ and $\bx\in E$ where $E$ is a compact subset of $\mathbb{R^d}$. 
The code which computes $f$ is expensive since the model is complex. 
We assume that no more than $N$ calls to $f$ are possible.
The input $\bx$ are measured with a lack of precision and some variables are uncontrollable.
Both sources of uncertainties are modeled by a random distribution on $E$. 
Let $\bX$ be the random variable. 
Our goal is to propose an upper bound of the probability: 

$$
\pi_\rho=\P(f(\bX)<\rho))=\P(\bX\in R_\rho)=\P_{\bX}(R_\rho)\,,
$$
where $R_\rho$ is a subset of $E$ defined by $R_\rho=\{\bx : f(\bx)< \rho\}$ and $\rho \in \mathbb{R}$
is a given threshold.

In a crude Monte Carlo scheme, the following estimator of $\pi_\rho$ is obtained:

\begin{equation}
 \label{estimateurMC}
\hat \pi_{\rho,N} = \frac{\Gamma(f,\bX_{1:N},\rho)}{N}\,, \\
\end{equation}
where $\Gamma(f,\bX_{1:N},\rho)$ is defined by

\begin{equation}
 \label{binomiale}
 \Gamma(f,\bX_{1:N},\rho)=\sum_{i=1}^N\I_{]-\infty,\rho[}(f(\bX_i))\,,\\
\end{equation}
and $\bX_{1:N}=(\bX_1,\ldots,\bX_N)$ is an $N$-sample of random variables
with the same distribution as $\bX$. Its expectation and its variance are:

$$
\E(\hat \pi_{\rho,N})=\mathbb{P}(\bX\in R_{\rho})=\pi_{\rho}\,, \quad \mathbb{V}(\hat \pi_{\rho,N})=\frac{1}{N}\pi_{\rho}(1-\pi_{\rho})\,.\\
$$
Since $\Gamma(f,\bX_{1:N},\rho)$ follows a binomial distribution with parameters $N$
and $\pi_\rho$, an exact confidence upper bound on $\pi_\rho$:

$$\P(\pi_\rho\le b(\Gamma(f,\bX_{1:N},\rho),N,\alpha))\ge1-\alpha\,,$$
is available. 

\newpage

\noindent Indeed, let $T$ be a random variable which follows a binomial distribution with parameters
$N$ and $p$. For any real number $\alpha\in[0,1]$, we can easily show that the upper confidence bound $b$ on $p$:

$$
\P_T(p\le  b(T,N,\alpha))\ge1-\alpha
$$
is such that:
\begin{equation}
\left\{
\begin{array}{ll}
 b=1                                                                                    & \text{if }T=N \\
 b\text{ is the solution of equation } \sum_{k=0}^T \binom{N}{k} b^k(1- b)^{N-k}=\alpha & \text{otherwise}
\end{array}
\right.\,.
\label{eq:binomial-bound}
\end{equation}
This upper bound is not in closed form but easily computable. 

In the case where $\Gamma(f,\bX_{1:N},\rho)=0$ which happens with probability $(1-\pi_{\rho})^N$, the 
$(1-\alpha)$-confidence interval is $[0,1-(\alpha)^{1/N}]$.
As an example, if the realization of $\Gamma(f,\bX_{1:N},\rho)$ is equal to $0$, 
an upper confidence bound at level $0.9$, $\pi_\rho\le 10^{-5}$ can be warranted only if
more than 230,000 calls to $f$ were performed. \\
When the purpose is to assess the reliability of a system under the constraint of a limited number of calls
to $f$, there is a need for a sharper upper bound on $\pi_\rho$.
Several ways to improve the precision of estimation and bounding have been proposed in the literature.

Since Monte Carlo estimation works better for frequent events,
the first idea is to change the crude scheme in such a manner that the event becomes less rare.
It is what importance sampling and splitting methods schemes try to achieve.\\
For example \citet{lecuyer:demers:tuffin:2007} showed that randomized quasi-Monte Carlo can be used jointly
with splitting and/or importance sampling.
By analysing a rare event as a cascade of intermediate less rare events,
\citet{delmoral:garnier:2005} developed a genealogical particle system approach to explore
the space of inputs $E$.
\citet{cerou:guyader:2007a,cerou:guyader:2007b} proposed an adaptive multilevel splitting also based
on particle systems.
An adaptive directional sampling method is presented by \citet{munoz:garnier:remy:deroquigny:2010} to accelerate 
the Monte Carlo simulation method. 
These methods can still need too many calls to $f$ and
the importance distribution is hard to set for an importance sampling method. 

A general approach in computer experiments is to make use of a metamodel
which is a fast computing function which approximates $f$.
It has to be built on the basis of data $\{f(\bx_1),\cdots,f(\bx_n)\}$
which are evaluations of $f$ at points of a well-chosen design $D_n=\{\bx_1,\cdots,\bx_n\}$. The bet is that
these $n$ evaluations will allow the building of more accurate bounds on the probability
of the target event.\\
Kriging is such a metamodeling tool: one can see \cite{santner:williams:notz:2003} and more recently
\citet{li:sudjianto:2005,joseph:2006,bingham:hengartner:higdon:kenny:2006}.
The function $f$ is seen as a realization of a Gaussian process which is a Bayesian prior. \\
The related posterior distribution is computed conditionally to the data.
It is still a Gaussian process whose mean can be used as a prediction of $f$ everywhere on $E$ 
and the variance as a pointwise measure of the confidence one can have in the prediction.\\
By using this mean and this variance, \citet{oakley:2004} has developed a sequential method 
to estimate quantiles and \citet{vazquez:bect:2009} a sequential method to estimate the probability
of a rare event.
\citet{cannamela:garnier:iooss:2008} have proposed some sampling strategies based only
on a reduced model which is a coarse approximation of $f$ 
(no information about the accuracy of prediction is given), to estimate quantiles.


In this paper, we also use Kriging metamodeling. Indeed, we assume that $f$ is a realization 
of a Gaussian process $F$. This Gaussian process is assumed independent of $\bX$ since it models the uncertainty in our knowledge of $f$ while $\bX$ models
a physical uncertainty on the input variables. 
As a consequence, $\pi_\rho$ is a realization of the random variable: 

$$\Pi_\rho=\E(\I_{]-\infty,\rho[}(F(\bX))|F)\,.$$
A natural approach consists of focusing on the posterior distribution of $\Pi_\rho$
which depends on the posterior distribution of $f$ given its computed evaluations. 
A Bayesian estimator of $\Pi_\rho$ can be computed 
and a credible bound is reachable by simulating realizations of the conditional Gaussian process
to obtain realizations of $\Pi_\rho$.\\
We propose another approach which makes use of an importance sampling method the importance distribution of which
is based on the metamodel.\\

The paper is organized as follows: Section \ref{secBayes} describes the
posterior distribution of the Gaussian process and how to obtain an estimator
and a credible bound of $\Pi_\rho$.
Section \ref{secIS} presents our importance sampling method and
the stochastic upper bound which is provided with a high probability.
Finally in Section \ref{secExample},
the two methods are compared on a toy example. Different designs of numerical experiments (sequential and non sequential) are performed for this comparison.
A solution
to a real aeronautical case study about the risk that the  trajectory of an airborne
load will collide with the aircraft that has released it, is proposed.

\section{Standard Bayesian bounds}
\label{secBayes}

The first step for Kriging metamodeling is to choose a design $D_n=\{\bx_1,\ldots,\bx_n\}$ of numerical experiments
(one can see \citet{morris:mitchell:1995,koehler:owen:1996} and more recently \cite{fang:li:sudjianto:2006,mease:bingham:2006,dette:pepelyshev:2010}).
Let $y_{D_n}=(y_1=f(\bx_1),\ldots, y_n=f(\bx_n))$ be the evaluations of $f$ on $D_n$.

\noindent Let us start from a statistical model consisting of Gaussian processes $F_{\bbeta,\sigma,\btheta}$,
the expressions of which are given by: for $\bx\in E$,
\begin{equation}
 \label{krigeage}
F_{\bbeta,\sigma,\btheta}(\bx)=\sum_{k=1}^L\beta_jh_j(\bx)+\zeta(\bx)=H(\bx)^T\bbeta+\zeta(\bx)\,,
\end{equation}
where 
\begin{itemize}
\item $h_1,\ldots,h_L$ are regression functions, 
and $\bbeta=(\beta_1,\ldots,\beta_L)$ is a vector of parameters,
\item $\zeta$ is a centered Gaussian process with covariance 
$$\text{Cov}(\zeta(\bx),\zeta(\bx'))=\sigma^2 K_{\btheta}(\bx,\bx')\,,$$ 
where $K_{\btheta}$ is a correlation function depending on some parameters $\btheta$
\citep[for details about kernels, see][]{koehler:owen:1996}. 
\end{itemize}
The maximum likelihood estimates $\hat \bbeta,\hat {\sigma}, \hat \btheta$ of $\bbeta,\sigma,\btheta$
are computed on the basis of the observations. Then, the Bayesian prior on $f$ is chosen to be
$F=F_{\hat \bbeta,\hat \sigma,\hat \btheta}$ and the process $F$ is assumed independent of $\bX$.
We denote $F^{D_n}$ the process $F$ conditionally to $F(\bx_1)=y_1,\ldots,F(\bx_n)=y_n$,
in short $Y_{D_n}=y_{D_n}$.


\noindent The process $F^{D_n}$ is still a Gaussian process \citep[see][]{santner:williams:notz:2003} with
\begin{itemize}
 \item mean: $\forall \bx$, 
\begin{equation}\label{mean}
m_{D_n}(\bx)=H(\bx)^T\hat\bbeta +\Sigma_{\bx {D_n}}^T\Sigma_{{D_n}{D_n}}^{-1}(y_{D_n}-H_{D_n}\hat\bbeta)\,, 
\end{equation}
\item covariance: $\forall \bx,\bx',$
\begin{equation}
K_{D_n}(\bx,\bx') = \hat\sigma^2(K_{\hat \btheta}(\bx,\bx')
-\Sigma_{\bx {D_n}}^T\Sigma_{{D_n}{D_n}}^{-1}\Sigma_{\bx' {D_n}})\,, 
\end{equation}
\end{itemize}
where 
$$
(\Sigma_{{D_n}{D_n}})_{1\leq i,j \leq n} =K_{\hat \btheta}(\bx_i,\bx_j) 
\text{ and } \Sigma_{\bx {D_n}}=\left(K_{\hat \btheta}(\bx,\bx_i)\right)_{1\le i \le n}^T\,.
$$
In this approach the conditioning to the data regards the parameters as fixed although they are estimated.

The Bayesian prior distribution $\P_F$ on $f$ leads to a Bayesian prior distribution on $\Pi_\rho$.
Our goal is to use the distribution of the posterior process $F^{D_n}$ conditionally to the observation
of $Y_{D_n}$, to learn about the posterior distribution of $\Pi_\rho$.
It is straightforward to show that $\Pi_\rho$ given $Y_{D_n}=y_{D_n}$, denoted $\Pi_\rho^{D_n}$,
has the same distribution as $\E(\I_{]-\infty,\rho[}(F^{D_n}(\bX))|F^{D_n})$.
The mean and the variance of $\Pi_\rho^{D_n}$ are then given by:
\begin{equation}
\label{postmean}
\E(\Pi_\rho^{D_n})=\int_{E}\E(\I_{]-\infty,\rho[}(F^{D_n}(\bx)))\P_{\bX}(d\bx)
=\E\left(\Phi\left(\frac{\rho-m_{D_n}(\bX)}{\sqrt{K_{D_n}(\bX,\bX)}}\right)\right)\,,
\end{equation}
where $\Phi$ is the cumulative distribution function of a centered reduced Gaussian random variable,

\begin{equation}
\label{postvariance}
\mathbb{V}(\Pi_\rho^{D_n})=
\int_{E\times E}\Covariance(\I_{]-\infty,\rho[}(F^{D_n}(\bx)),\I_{]-\infty,\rho[}(F^{D_n}(\bx'))
\P_{\bX}\times \P_{\bX}(d\bx,d\bx')\,.
\end{equation}
%
%


A numerical Monte Carlo integration can be used to compute the posterior mean and variance
since they do not need more calls to $f$.
However, the  computation time requested by a massive Monte Carlo integration, 
especially for $\mathbb{V}(\Pi_\rho^{D_n})$, can be very long as can be seen in the examples.

The mean and the variance of $\Pi_\rho^{D_n}$ can be used to obtain credible bounds.
As a consequence of Markov inequality, it holds, for any $\alpha\in[0,1]$,
\begin{equation}
\label{markovbound}
\P\left(\Pi_\rho^{D_n} \le \frac{\E(\Pi_\rho^{D_n})}{\alpha}\right)\ge 1-\alpha\,.
\end{equation}
Likewise, Chebychev inequality gives, for any $\alpha\in[0,1]$,
\begin{equation}
\label{chebychevbound}
\P\left(\Pi_\rho^{D_n}\le \E(\Pi_\rho^{D_n}) + 
\sqrt{\frac{\mathbb{V}(\Pi_\rho^{D_n})}{\alpha}}\right)\ge 1-\alpha\,.
\end{equation}

Moreover, the quantiles of $\Pi_\rho^{D_n}$ can be estimated through massive simulations of the conditional process $F^{D_n}$. 
These realizations of $F^{D_n}$ lead to realizations of $\Pi_\rho^{D_n}$ from which quantiles can be estimated.
These quantiles of $\Pi_\rho^{D_n}$ are exactly the upper bounds that are sought.  

We adapt the algorithm proposed by \citet{oakley:2004} to obtain realizations of $\Pi_\rho^{D_n}$.
From a realization $F^{D_n}$, the corresponding realization of $\Pi_\rho^{D_n}$ is computed using a massive Monte Carlo integration 
with respect to the distribution of $\bX$. 
Thus, a credible bound on $\Pi_\rho^{D_n}$ is constructed. 
Given $\alpha\in (0,1)$, a constant $a\in[0,1]$ is found such that:

$$
\P(\Pi_\rho^{D_n}<a)\ge 1-\alpha\,.
$$

However, it is not possible to get an exact realization of $F^{D_n}$ or 
to sample jointly $F^{D_n}$ at all the inputs of the Monte Carlo sample (of $\bX$)
since this sample is too large.
Thus, the algorithm relies on a discretization of the process. The same scheme as the one followed by
\citet{oakley:2004} is used. 
We choose $T$ points in $E$: $D'=\{\bx_1',\ldots,\bx_T'\}$ where the corresponding realizations of $F^{D_n}$
are simulated. From the set $\{y_1',\ldots,y_T'\}$ of joint realizations of $\{F^{D_n}(\bx_1'),\ldots,F^{D_n}(\bx_T')\}$, 
a realization of $F^{D_n}$ is approximated by the mean of $F^{D_n,D'}$ which is the process $F$ conditioned to $F(\bx_1)=y_1,\ldots,F(\bx_n)=y_n$ and
$F(\bx_1')=y_1',\ldots,F(\bx_T')=y_T'$. 
The variance of $F^{D_n,D'}$ has to be very small for any point in $E$ for the approximation to be valid. 
Hence, $T$ has to be large enough and the points in $D'$ have to fill the space $E$.
When the dimension of the input space is low, to propose such a set $D'$ is quite easy. However,
it can be burdensome to fill the space in high dimension and it can lead to a too large number
of needed realizations of $F^{D_n}$ which are impossible to simulate jointly. 
The discretization step is a major concern since it induces an uncontrollable error on the credible bound on $\Pi_\rho^{D_n}$.
We propose then in the next Section, an alternative approach based on importance sampling which avoids this discretization step.

\section{Metamodel-based importance sampling}
\label{secIS}
As was explained in Section \ref{secintroduction}, the major drawback of the crude Monte Carlo
scheme is the high level of uncertainty when it is used for estimating
the probability of a rare event.
Importance sampling is a way to tackle this problem. 
The basic idea is to change the distribution to make the target event more frequent.
We aim at sampling according to the importance distribution:

$$\P_{\bZ} : A\subset E \mapsto \P_{\bX}(A|\hat R_\rho)\,,$$
where $\hat R_\rho\subset E$ is to be designed close to $R_\rho=\{\bx\in E : f(\bx)<\rho\}$.
Thanks to $n$ calls to the metamodel, a set $\hat R_{\rho}$ can be chosen as follows:
\begin{equation}
\label{rchap} 
\hat R_{\rho}=\hat R_{\rho,\kappa}=\left\{\bx : m_{D_n}(\bx) <\rho+\kappa\sqrt{K_{D_n}(\bx,\bx)}\right\}\,,
\end{equation}
where $\kappa$ is fixed such that ``$\{\bx: F(\bx)<\rho\}\subset\hat R_{\rho,\kappa}$ with a good confidence level''. 
In other words, if $\bx$ is such that $f(\bx)<\rho$, we want $\bx$ to be in $\hat R_{\rho,\kappa}$.
We recall that the posterior mean $ m_{D_n}(\bx)$ is an approximation of $f(\bx)$ and $\kappa\sqrt{K_{D_n}(\bx,\bx)}$
has been added to take into account the uncertainty of the approximation.

A set of $m$ points, $\bZ_{1:m}=(\bZ_1,\ldots,\bZ_m)$, is drawn to be an i.i.d. sample following the 
importance distribution.
The corresponding importance sampling estimator of $\pi_\rho$ is 
\begin{equation}
\label{isestimator}
\frac{\P_{\bX}(\hat R_{\rho})}{m}\Gamma(f,\bZ_{1:m})=\frac{\P_{\bX}(\hat R_{\rho})}{m}\sum_{k=1}^m 
\I_{]-\infty,\rho[}(f(\bZ_k))\,.
\end{equation}

The probability $\P_{\bX}(\hat R_{\rho})$ is computable by a Monte Carlo integration since
it does not depend on $f$; yet, $m$ more calls to $f$ are necessary to compute $\I_{]-\infty,\rho[}(f(\bZ_k))$.
This estimator is only unbiased provided that $R_{\rho}\subset\hat R_{\rho}$.
Nevertheless, it is an unbiased estimator of
$\E_{\bX}(\I_{]-\infty,\rho[}(f(\bX))\I_{\hat R_{\rho}}(\bX))$.
Since $\Gamma(f,\bZ_{1:m})$ follows a binomial distribution \\
$\mathcal{B}\left(m,\frac{\E(\I_{]-\infty,\rho[}(f(\bX))\I_{\hat R_\rho}(\bX))}{\P_{\bX}(\hat
 R_\rho)}\right)$, for any $\alpha\in]0;1[$, the following confidence upper bound holds:
\begin{equation}
\label{boundisf}
\P\left(\E(\I_{]-\infty,\rho[}(f(\bX))\I_{\hat R_\rho}(\bX))\le 
b(\Gamma(f,\bZ_{1:m},\rho),m,\alpha)\P_{\bX}(\hat R_\rho)\right)>1-\alpha \,,
\end{equation}
by using the bound (\ref{eq:binomial-bound}).
This is an upper bound on $\pi_\rho$ only if the estimator (\ref{isestimator}) is unbiased i.e. only if
 $R_{\rho}\subset\hat R_{\rho}$.
As is noticed in the decomposition:

\begin {equation*}
\label{pirhodecomp}
\pi_\rho = \E(\I_{]-\infty,\rho[}(f(\bX)))=\E(\I_{]-\infty,\rho[}(f(\bX))\I_{\hat R_\rho}(\bX)) 
+ \E(\I_{]-\infty,\rho[}(f(\bX))(1-\I_{\hat R_\rho}(\bX)))\,,
\end{equation*}
the second term on the right-hand side which is the opposite of the bias has to be controlled.
That is why the random variable

$$\Pi_\rho^{D_n}=\E(\I_{]-\infty,\rho[}(F^{D_n}(\bX))|F^{D_n})\,,$$
whose a realization is $\pi_\rho$, is considered.\\
Similarly to the previous decomposition, it holds that
\begin{equation}
\label{Pirhodecomp}
\Pi_\rho^{D_n} =\E(\I_{]-\infty,\rho[}(F^{D_n}(\bX))\I_{\hat R_\rho}(\bX)|F^{D_n}) + \E(\I_{]-\infty,\rho[}(F^{D_n}(\bX))(1-\I_{\hat R_\rho}(\bX))|F^{D_n})\,.
\end{equation}
A bound on $\E(\I_{]-\infty,\rho[}(F^{D_n}(\bX))\I_{\hat R_\rho}(\bX)|F^{D_n})$ comes from
(\ref{boundisf}).
\begin{proposition}
\label{propborneis}
For $\alpha\in ]0,1[$, it holds that
\begin{equation}
\label{boundISF}  
\P(\left(\E(\I_{]-\infty,\rho[}(F^{D_n}(\bX))\I_{\hat R_\rho}(\bX)|F^{D_n})\le \bb\ \P_{\bX}(\hat R_\rho)\right)\ge 1-\alpha\,,
\end{equation}
where $\bb$ stands for $b(\Gamma(F^{D_n},\bZ_{1:m},\rho),m,\alpha)$.
\end{proposition}
\Preuve
Let $\varphi$ be any realization of $F^{D_n}$.\\
As in (\ref{boundisf}), we have

$$\P\left(\E(\I_{]-\infty,\rho[}(\varphi(\bX))\I_{\hat R_\rho}(\bX))\le b(\Gamma(\varphi,\bZ_{1:m},\rho),m,\alpha) \P_{\bX}(\hat R_\rho)\right)\ge 1-\alpha\,.$$
Thus, since this result holds for any realization of $F^{D_n}$,

$$\P\left(\E(\I_{]-\infty,\rho[}(F^{D_n}(\bX))\I_{\hat R_\rho}(\bX)|F^{D_n})\le b(\Gamma(F^{D_n},\bZ_{1:m},\rho),m,\alpha)\P_{\bX}(\hat R_\rho)\right)\ge 1-\alpha\,. $$
\Fin
The next proposition states an upper bound for the second term in (\ref{Pirhodecomp}).
\begin{proposition}
\label{propbiais}
For $\beta\in ]0,1[$, it holds that
$$\P\left(
\E(\I_{]-\infty,\rho[}(F^{D_n}(\bX))(1-\I_{\hat R_\rho}(\bX))|F^{D_n})\le
 \frac{\mathbf{c}}{\beta}\right)\ge 1- \beta\,,$$
where $\mathbf{c}=\E\left( \Phi\left(\frac{\rho-m_{D_n}(\bX)}{\sqrt{K_{D_n}(\bX,\bX)}}\right)(1-\I_{\hat R_\rho}(\bX))\right)$.
\end{proposition}
\Preuve
The mean of $\E(\I_{]-\infty,\rho[}(F^{D_n}(\bX))(1-\I_{\hat R_\rho}(\bX))|F^{D_n})$ can be computed
in the same fashion as the mean of $\Pi_\rho^{D_n}$.
It gives
{\small $$\E\left(\E(\I_{]-\infty,\rho[}(F^{D_n}(\bX))(1-\I_{\hat R_\rho}(\bX))|F^{D_n})\right)
=\E\left(\Phi\left(\frac{\rho-m_{D_n}(\bX)}{\sqrt{K_{D_n}(\bX,\bX)}}\right)(1-\I_{{\hat R_\rho}}(\bX))\right)\,.
$$}
Then, Markov inequality is applied which completes the proof.
\Fin
Finally, by gathering the results of Proposition \ref{propborneis} and Proposition \ref{propbiais}, a
stochastic upper bound is found on $\Pi_\rho^{D_n}$.
\begin{proposition}
\label{finalbound}
For $\alpha,\beta\in ]0,1[$ such that $\alpha+\beta<1$, it holds that
\begin{equation}
\label{finalboundeq}
\P\left(\Pi_\rho^{D_n}\le \bb \P_{\bX}(\hat R_\rho)+\frac{\mathbf{c}}{\beta}\right)\ge 1-(\alpha+\beta)\,,
\end{equation}
where $\mathbf{b}$ and $\mathbf{c}$ have been defined above.
\end{proposition}
The proof is obvious.

If $\hat R_\rho$ is chosen as proposed in (\ref{rchap}), the bound $\mathbf{c}$ is:
$$
\mathbf{c}=c(\kappa)=\E\left( \Phi\left(\frac{\rho-m_{D_n}(\bX)}
{\sqrt{K_{D_n}(\bX,\bX)}}\right)\I_{]-\infty,-\kappa[}\left(\frac{\rho-m_{D_n}(\bX)}{\sqrt{K_{D_n}(\bX,\bX)}}\right)\right)\,.
$$


\section{Numerical experiments}
\label{secExample}
\subsection{A toy example}
\label{evraretoyex}

We study on a toy example, described below, the two bounding strategies: the Bayesian strategy (credible bound obtained by simulating realizations of $F^{D_n}$
as described in the end of Section \ref{secBayes})
and the MBIS (metamodel-based importance sampling) strategy (stochastic bound given by Proposition \ref{finalbound}).
Since the credible bounds in the Bayesian strategy are directly derived from the metamodeling, the Bayesian strategy is the reference and if the dimension of $E$ is low,
it should perform well. Our aim is to test whether the MBIS strategy can achieve such good bounds. Since the choice in the 
design $D_n$ should directly impact the set $\hat R_{\rho,\kappa}$ (\ref{rchap}) and hence the quality of the importance sampling, 
different kind of designs are considered to compare the strategies.\\

\noindent The function $f:E=[-10,10]^2\rightarrow \mathbb{R}_+$ is assumed to describe a physical model:

$$
f(x_1,x_2)=-\frac{\sin(x_1)}{x_1}-\frac{\sin(x_2+2)}{x_2+2}+2\,.
$$

\begin{figure}[ht]
\begin{center}
\includegraphics[height=.4\textwidth]{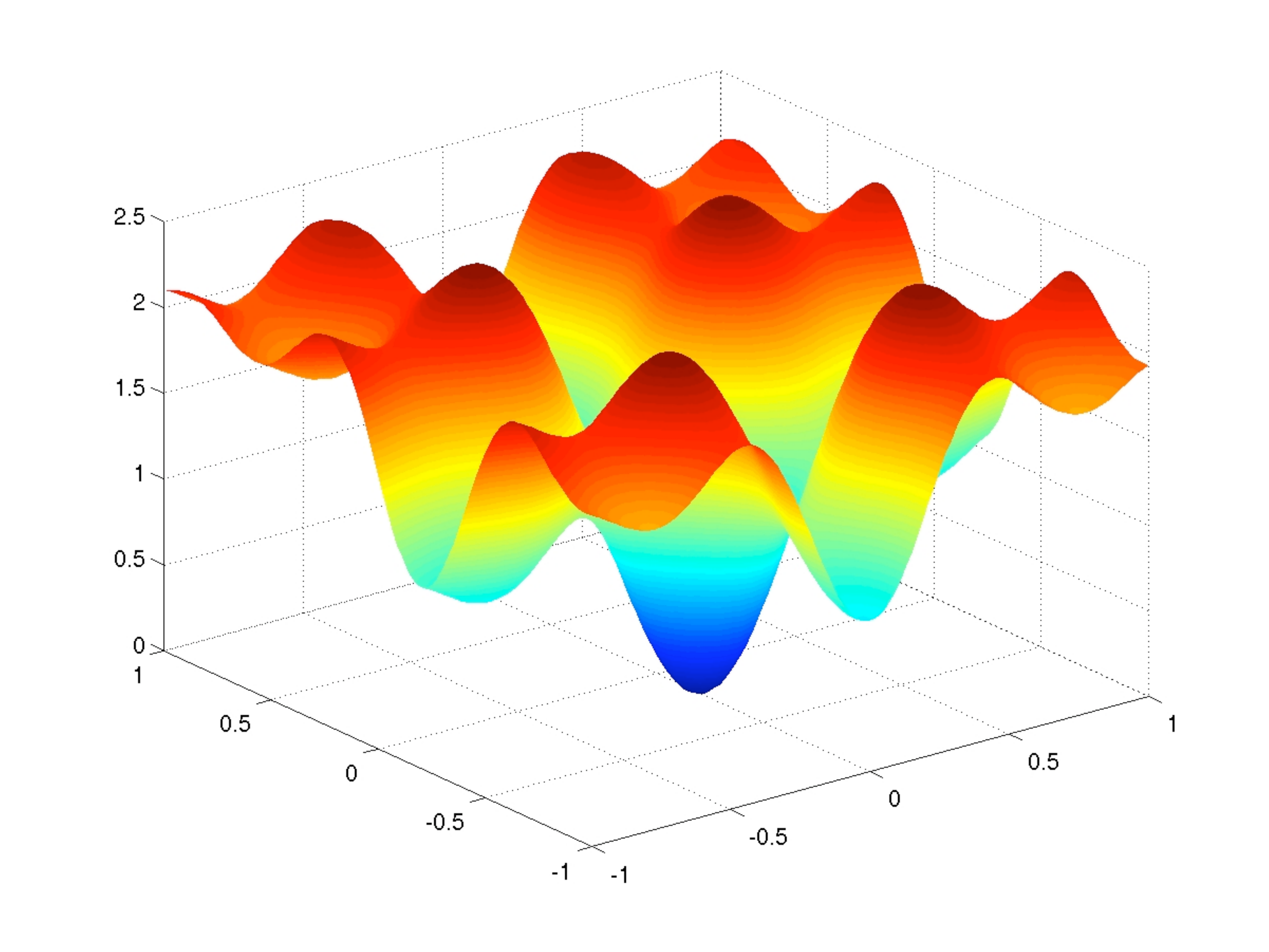}
\end{center}
\caption{The function $f$}
\end{figure}

The input vector $\bX$ is supposed to have a uniform distribution on $E$.
The threshold is set to $\rho=0.01$ which corresponds to the probability \\
$\P_{\bX}\left(f(\bX)<\rho\right)=4.72\cdot 10^{-4}$. This probability was computed thanks to a massive
Monte Carlo integration.\\
It is assumed that no more than $N=100$ calls to $f$ are allowed. 
For the Bayesian strategy, all of the $N=100$ available calls to $f$ are used to build the metamodel, while for the MBIS
strategy (using notations of Section \ref{secIS}) $n=50$ and $m=50$ are set.   
The two strategies are compared with different design sampling methods.
Three design sampling methods are used: an LHS-maximin method \citep{morris:mitchell:1995} which is non sequential 
and space filling and two sequential methods. The sequential sampling methods are based on a first LHS-maximin design including 
$80\%$ of the points and the last $20\%$ are added sequentially according to the criterion
tIMSE (targeted Integrated Mean Square Error) proposed by \citet{picheny:ginsbourger:roustant:haftka:2008} for
one method and according to the criterion J SUR (Stepwise Uncertainty Reduction)
proposed by \citet{bect:ginsbourger:li:picheny:vasquez:2011} for the other method. These sequential methods are based on
a trade-off between reduction of uncertainty in the knowledge of $f$ and the exploration of the space around the critical set $R_\rho$.
All Kriging metamodels are built with an intercept as the regression function and a Gaussian correlation function is chosen
as the correlation function of the Gaussian process $\zeta$ i.e. $\forall \bx\in E$, $h(\bx)=1$ and $\forall \bx,\bx'\in E$,
$K(\bx,\bx)=\exp\left(-\theta\Vert\bx-\bx'\Vert^2\right)$ are set for the model given by equation (\ref{krigeage}).
For the Bayesian strategy,
a thousand realizations of $\Pi_\rho^{D_N}$ are computed from which the credible bound is obtained. 
The discretization is done on a grid with $100$ points and to prevent ill-conditioned covariance matrices, 
if a point of the grid is too close to a point of the design $D_N$ it is replaced with a point in $E$ far enough from the points of the design and the points of the grid. 
The numerical integration to compute the realization of $\Pi_\rho^{D_N}$ from the realization of the process is done with a $10^5$-sample. 
In the MBIS strategy, the probability $\P_{\bX}(\hat R_{\rho,\kappa})$
(and also the bound on the bias, given in Proposition \ref{propbiais}) was computed by a Monte Carlo integration on
a $10^7$-sample and $\kappa=3$ has been set. 
\\

There are sources of variability on the estimators and the bounds due to the design sampling methods.
Indeed, the designs are computed to be maximin by using a finite number of iterations of a simulated annealing algorithm.
Moreover, there exist symmetries within the class of maximin designs. In the sequential method, the point to be added is sought
through a stochastic algorithm.
Concerning the importance sampling strategy, the sampling which gives $\bZ_{1:m}$ induces variability.
In order to test the sensitivity to these sources of variability,
each of the two strategies for each of the three design sampling methods is repeated one hundred times.

\begin{figure}[h!]
\begin{center}
\includegraphics[scale=0.3]{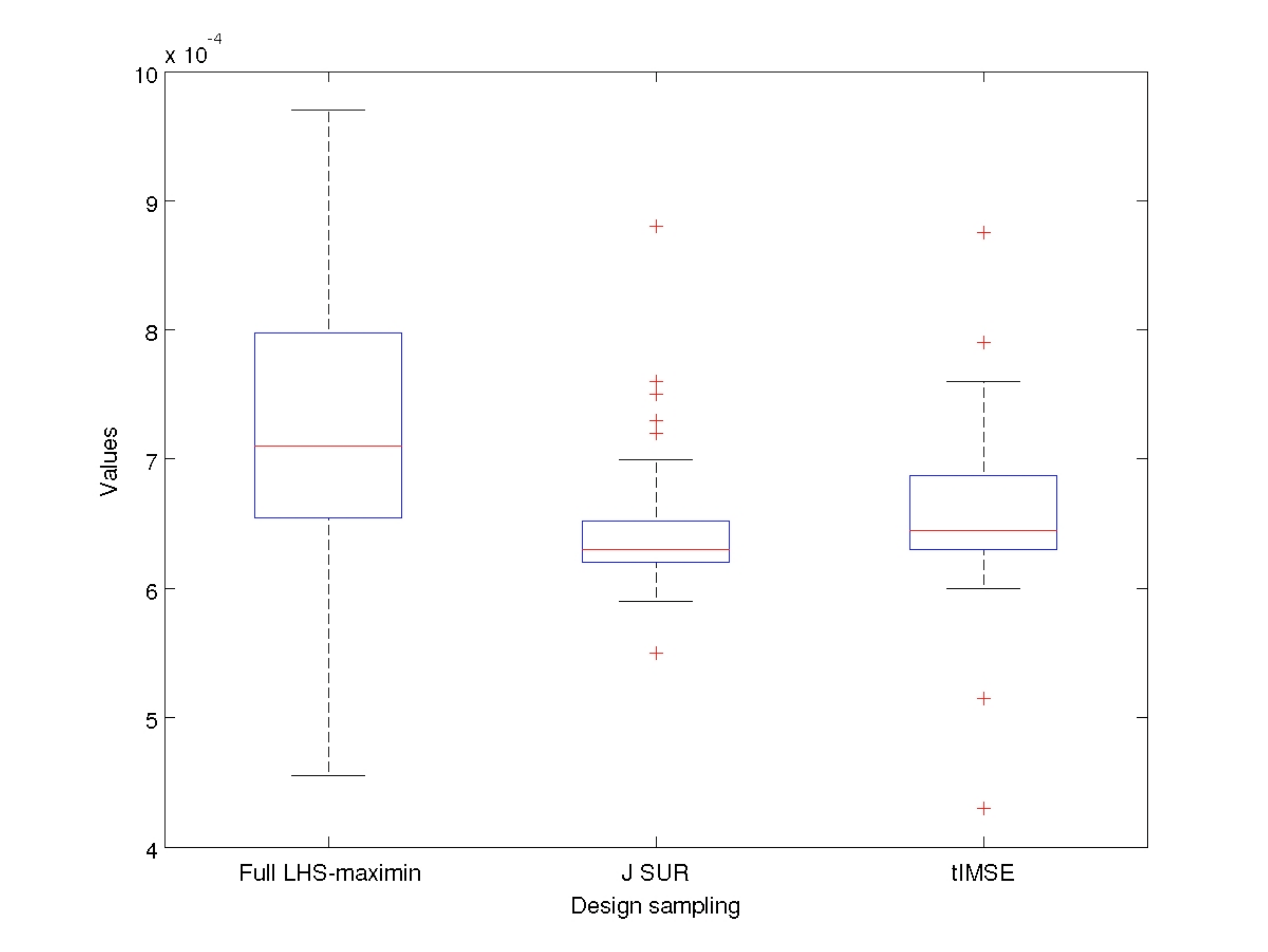}
\end{center}
\caption{Bayesian $98\%$ credible bound for $\pi_\rho$}
\label{boxplotBa}
\end{figure}

\begin{figure}[h!]
\begin{center}
\includegraphics[scale=0.3]{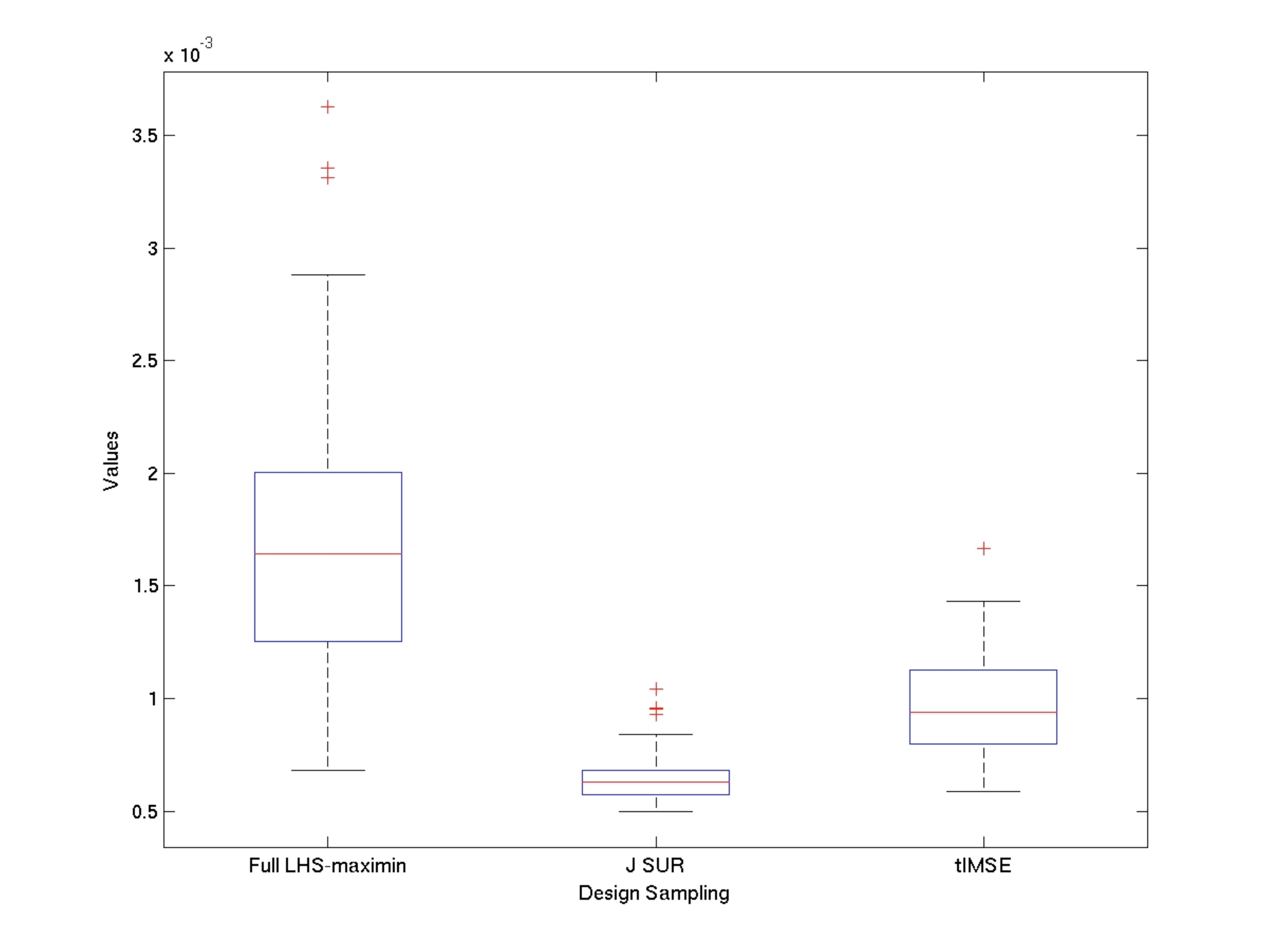}
\end{center}
\caption{MBIS $98\%$ stochastic bounds of $\pi_\rho$}
\label{boxplotIS}
\end{figure}

\begin{table}[h!]
\begin{center}
\begin{tabular}{|l|c|c|c|}
 \hline
& Full LHS-maximin & J SUR & tIMSE\\
\hline
Minimum& $4.55$&$5.50$&$4.30 $\\
\hline
$1^{\text{st}}$ quartile&$ 6.55$& $6.20 $&$ 6.30$ \\
\hline
Mean &$7.78$&$ 7.58$&$ 52.2$\\
\hline
Median &$7.10$&$ 6.30$&$ 6.45$\\
\hline
$3^{\text{rd}}$ quartile& $7.97$&$ 6.52$&$ 6.87$ \\
\hline 
Maximum& $35.3$&$55.4 $&$3027 $\\
\hline
\end{tabular}
\end{center}
\caption{Bayesian $98\%$ credible bound of $\pi_\rho$ multiplied by $10^{4}$}
\label{tableBabounds}
\end{table}

\begin{table}[h!]
\begin{center}
\begin{tabular}{|l|c|c|c|}
 \hline
& Full LHS-maximin & J SUR & tIMSE\\
\hline
Minimum& $6.82$&$4.98$&$5.88$\\
\hline
$1^{\text{st}}$ quartile &$12.58 $&$5.71$&$ 8.00$ \\
\hline
Mean &$16.8$&$6.43$&$9.67$\\
\hline
Median &$16.4 $&$ 6.28$&$9.39 $\\
\hline
$3^{\text{rd}}$ quartile &$20.1 $&$ 6.81$&$11.3$\\
\hline
Maximum& $36.3$&$10.4$&$16.7$\\
\hline
\end{tabular}
\end{center}
\caption{IS $98\%$ stochastic bounds of $\pi_\rho$ multiplied by $10^{4}$}
\label{tableISbounds}
\end{table}

Figure \ref{boxplotBa} and Table \ref{tableBabounds} display the results for $98\%$ credible bounds obtained by the Bayesian strategy and 
Figure \ref{boxplotIS} and Table \ref{tableISbounds} display the results for $98\%$ stochastic bounds obtained by the MBIS strategy.
The results are provided according to the design sampling method.
For the MBIS stochastic bounds, $\alpha=1\%$ and $\beta=1\%$ have been set using the notations of Proposition \ref{finalbound}. 
If a crude Monte Carlo scheme as presented in Section \ref{secintroduction} is used here with only $N=100$ calls, the estimator is equal to $0$ with probability greater than 
$0.95$ and in this case, the upper confidence bound is $0.038$ at level $98\%$.
The two strategies bring much sharper bounds on the probability of the rare event. 
The sequential design sampling method with J SUR criterion leads to the better bounds whatever the strategy. 
The MBIS strategy manages to reach the same sharp bounds as the ones provided by the Bayesian strategy in the case
where the design is obtained thanks to the J SUR criterion.
The Bayesian strategy is less sensitive 
to the choice in the design sampling method. However, the Bayesian method suffers
from the fact that the quantiles are estimated thanks to conditional simulations of the Gaussian process which rely
on a discretization of the space. Hence, it leads to an approximation in results and stability concern as is noticed in the results (see the 
maximum credible bound obtained with the
tIMSE sampling criterion).
Furthermore, a limited number of iterations is achievable since the simulations are quite burdensome.

The bound provided by Markov inequality (\ref{markovbound}) is not interesting in this example since
it cannot be less than 0.01. Chebychev inequality has not been used since we were not able to determine the posterior variance in a reasonable time. 
\\

As the strategies depend on the Kriging model hypothesis (\ref{krigeage}),
a leave-one-out cross validation as proposed by \citet{jones:schonlau:welch:1998} can be performed
to check whether this hypothesis is sensible. It consists of building $n$ metamodels with posterior
mean and variance denoted respectively by  $m_{D_n^{-i}}$ and $\sigma^2_{D_n^{-i}}$, from designs 

$$
D_n^{-i}=\{\bx_1,\ldots,\bx_{i-1},\bx_{i+1},\ldots,\bx_n\}\,,
$$
where $i=1,\ldots,n$. \\
Then, the values 
\begin{equation}
\frac{|f(\bx_i)-m_{D_n^{-i}}(\bx_i)|}{\sigma^2_{D_n^{-i}}(\bx_i)}\,,
\end{equation}
are computed.
If something like $99.7\%$ of them lie in the interval $[-3,3]$, the Kriging hypothesis is not rejected.
In our toy example, in all of the tests which were done, all these values are in $[-2,2]$.


\subsection{A real case study: release envelope clearance}

\subsubsection{Context}
When releasing an airborne load, a critical issue is the risk that its trajectory could collide with the aircraft.
The behavior of such a load after release depends on many variables. Some are under the control of the crew: mach,
altitude, load factor, etc. We call them \emph{controlled variables} and denote their variation domain as $C$.
The others are \emph{uncontrolled variables}: let $E$ be the set of their possible values.
The \emph{release envelope clearance} problem consists of exploring the set $C$ to find a
subset where the release is safe, whatever the uncontrolled variables are.
To investigate this problem, we can use a simulator which computes the trajectory of the carriage when the
values of all the variables are given. Moreover, for $\bx_{C}\in C$ and $\bx\in E$, besides the trajectory
$\tau(\bx_{C},\bx)$, the program delivers a \emph{danger score} $f(\bx_{C},\bx)$ to be interpreted
as an ``algebraic distance'': a negative value characterizes a collision trajectory. \\
To assess the safety of release at a given point of $C$, we suppose that the values of the uncontrolled
variables are realizations of a random variable $\bX \in E$ that can be simulated.
Therefore, for a given value $\bx_{C}\in C$, and $\rho\ge 0$ the $\rho$-collision risk is the probability

$$
\pi_\rho(\bx_{C})= \P(f(\bx_{C},\bX) <\rho)\,.
$$

We do not aim at estimating this risk accurately.\\
We would rather classify the points into three categories: according to the position of $0$-risk $\pi_0(\bx_{C})$ with
respect to the two markers $10^{-5}$ and $10^{-2}$, $\bx_{C}$ is said to be
\begin{enumerate}
\item   {\bf totally safe  }  if $\pi_0(\bx_{C})\le 10^{-5}$,
\item {\bf relatively safe  } if $10^{-5}<\pi_0(\bx_{C})<  10^{-2}$,
\item  {\bf unsafe   } if 
$\pi_0(\bx_{C})\ge 10^{-2}$.
\end{enumerate}

In this example, there are $5$ controlled and $26$ uncontrolled variables, so that $C\subset \mathbb{R}^5, E \subset \mathbb{R}^{26}$.
From budgetary point of view, experts consider that a set of about $400$ representative points of $C$ 
is enough to cover the domain $C$ consistently.
On the other hand, the computation of $800,000$ trajectories takes about 4 days which is considered
reasonable.
On the basis of these indications, the maximum amount of available calls to the simulator 
is $N=2000$ per point.

\subsubsection{Bounding strategy}
\label{strategy}

As the dimension of the set of uncontrolled variables $E$ is high, the credible bounds obtained with the Bayesian strategy 
are impossible to get. The stochastic bounds provided by the MBIS strategy are still available. Unfortunately, 
a sequential sampling method for the design of experiments is not achievable since the simulator is too expensive if only one point is evaluated per call. 
Indeed, a fixed part of the cost of a call does not depend on the number of points for which the code is run. 
Although the MBIS strategy with an LHS-maximin sampling method is not optimal, it is still efficient.
\\    

We propose this two-step bounding strategy which can be applied for each point of the set of representative points (in the set $C$). 
Each step uses half of the calls budget: $m=n=\frac{N}{2}=1000$.
Let $\bx_{C}\in C$ be the current point of interest that we suppose fixed. For any $\bx\in E$,
$f(\bx) = f(\bx_{C},\bx)$ is set. It then corresponds to the notation introduced in the first part of the paper.
\begin{enumerate}
\item At the first stage, a Gaussian process is built as explained in (\ref{secBayes}), on the basis of evaluations
$f(\bx_1),\cdots,f(\bx_n)\in \mathbb{R}^n$ of $f$ on $D_n = \{\bx_1,\cdots,\bx_n\}$.
We know that $\pi_\rho$ is a realization of the random variable $\Pi_\rho^{D_n}$
whose mean

$$
\E(\Pi_\rho^{D_n})=\E\left(\Phi\left(\frac{\rho-m_{D_n}(\bX)}{\sqrt{K_{D_n}(\bX,\bX)}}\right)\right)\,,
$$
can be computed  accurately.\\
As stated by (\ref{markovbound}), applying Markov inequality gives, for any $\alpha\in ]0;1[$,

$$\P\left(\Pi_\rho^{D_n}\le \frac{\E(\Pi_\rho^{D_n})}{\alpha}\right)\ge 1-\alpha\,.$$
According to the value of $\E(\Pi_\rho^{D_n})$ we then take the following decisions:
\begin{itemize}
\item if $\E(\Pi_\rho^{D_n})\le \frac{1}{2}10^{-10}$ which leads by (\ref{markovbound}) to
$\P\left(\Pi_\rho^{D_n}\le \frac{10^{-5}}{2}\right)\ge 1-\frac{10^{-5}}{2}$, we qualify the current point $\bx_{C}\in C$ as totally safe,
\item if $\E(\Pi_\rho^{D_n})\ge 10^{-2}$, we conservatively classify $\bx_{C}$ as unsafe,
\item if $\frac{1}{2}10^{-10}<\E(\Pi_\rho^{D_n})< 10^{-2}$ we use a second stage procedure to refine the risk assessment.
\end{itemize}
\item A  million-sample $\bx_1,\cdots,\bx_M$ of $\bX$ is drawn from which we tune $\kappa$ in such a way that $m=1000$
of these million elements of $E$ are in $\hat R_{\rho,\kappa}$. The resulting points $\bz_1,\cdots,\bz_m$ are an $m$-sample
$\bz_{1:m}$ of realizations of the random variable $\bZ$ which follows
the importance distribution,\\

$$\P_{\bZ}:A\mapsto \P_{\bX}(A|\hat R_{\rho,\kappa})\,.$$
By using $m$ calls to the simulator, $\Gamma(f,\bz_{1:m},\rho)$ is computed.
Drawn from Proposition \ref{finalbound} with setting $\alpha=\beta$, we obtain the bound

$$ b(\Gamma(f,\bz_{1:m},\rho),m,\alpha)\P_{\bX}(\hat R_{\rho,\kappa})+\frac{c(\kappa)}{\alpha}\,,$$
which is a decreasing function of $\alpha$.\\
Let us define $\alpha_0 = \min \{\alpha : b(\Gamma(f,\bz_{1:m},\rho),m,\alpha)\P_{\bX}(\hat R_{\rho,\kappa})+\frac{c(\kappa)}{\alpha}\le 2\alpha\}$.
For such an $\alpha_0$, Proposition \ref{finalbound}
states:

$$
\P\left(\Pi_\rho^{D_n}\le b(\Gamma(F^{D_n},\bZ_{1:m},\rho),m,\alpha_0)\P_{\bX}(\hat R_\rho)+\frac{c(\kappa)}{\alpha_0}\right)\ge 1-2\alpha_0\,,
$$
which provides $2\alpha_0$ as a $1-2\alpha_0$ confidence upper bound on $\pi_\rho$.
\end{enumerate}

\subsubsection{Experiments}  
\begin{figure}[htbp]
\begin{center}
\includegraphics[height=.7\textwidth]{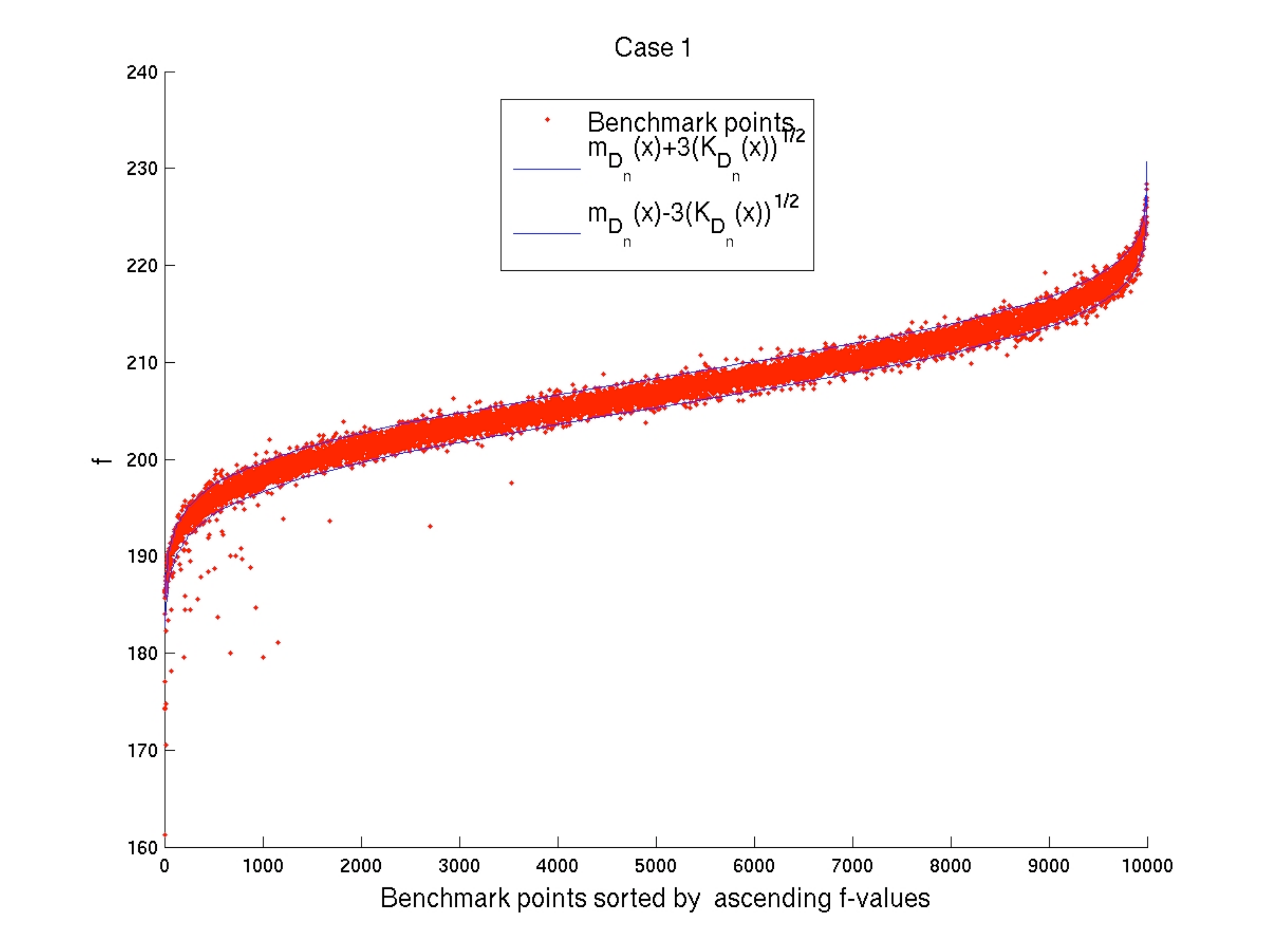}
\end{center}
\caption{Prediction performance case 1}
\label{case1}
\end{figure}

\begin{figure}[htbp]
\begin{center}
\includegraphics[height=.7\textwidth]{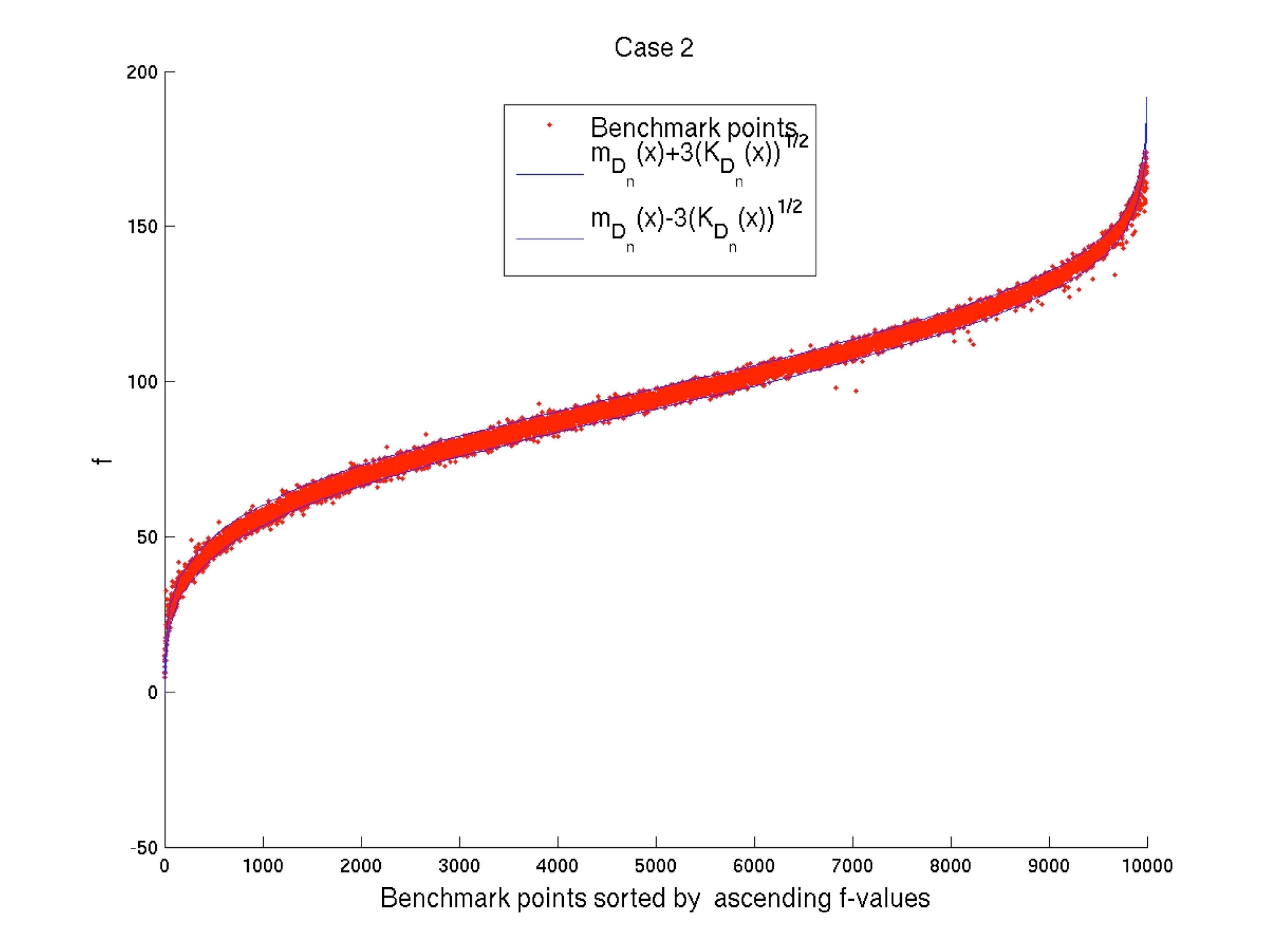}
\end{center}
\caption{Prediction performance case 2}
\label{case2}
\end{figure}

\begin{figure}[htbp]
\begin{center}
\includegraphics[height=.7\textwidth]{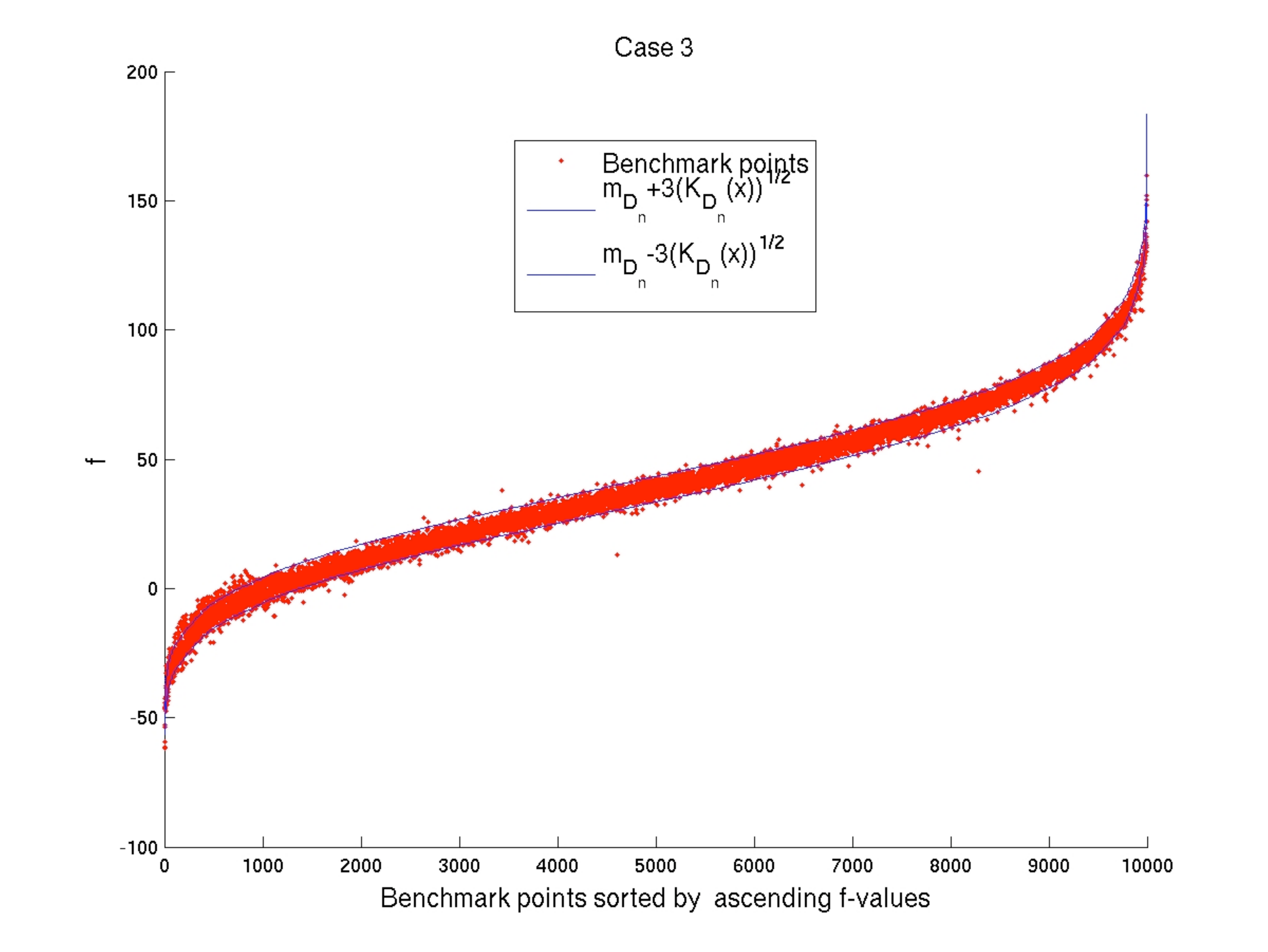}
\end{center}
\caption{Prediction performance case 3}
\label{case3}
\end{figure}

Three points of $C$ have been experienced.
Of these cases the first one is known to be a null $0$-risk point, while the third one is very unsafe and
the second one is in-between.
For benchmarking purposes, besides the simulator calls budget required for the estimation process described in \ref{strategy},
a $10,000$ sample of realizations of $f(\bx_{E},\bX)$ has been computed for each of the three examples.
For each case, we began by estimating a Gaussian process on the basis of $ f$-values computed on the points of a 
$1000$-point design $D_n=\{\bx_1,\cdots,\bx_n\}$.
This design was obtained by LHS-maximin sampling. 
Figures \ref{case1}, \ref{case2} and \ref{case3} show the predictive performance of the processes when applied to the benchmark points.  
These points, which appear in red, are sorted according to their process mean values while the blue curves
mark the predicted 3 standard deviation positions around the means. As appears rather clearly, the dispersion of the
real values is underestimated by the model: they overflow the blue zone with a frequency ($\sim 5\%$) higher
than expected ($0.27\%$).
The worst case is the first one, for which large deviations appear for benchmark points with low values of $f$.
In order to obtain  bounds from (\ref{markovbound}), we then computed $\E(\Pi_0^{D_n})$ using (\ref{postmean}):
\begin{itemize}
\item In the first case, the massive Monte Carlo procedure leads to a numerically null evaluation of $\E(\Pi_0^{D_n})$ 
and, as a consequence, to the classification of the related $C$ point as totally safe.
\item In the second example, $\E(\Pi_0^{D_n})$ being evaluated at $1.68\ 10^{-4}$, we need to proceed to the second step
of the bounding strategy to refine the collision probability estimation. 
The obtained 
confidence upper bound is $1.2\ 10^{-5}$ at confidence level $1-1.2\ 10^{-5}$.
The benchmark data do not show collision case: a $90\%$ confidence upper bound is $2.3\ 10^{-4}$. 
\item $\E(\Pi_0^{D_n})=0.103$ in case 3 which is consistent with the $90\%$ confidence interval $[0.0999;0.1101]$,
obtained on benchmark data.
\end{itemize}

\section{Discussion}

In this paper, we have especially focused on bounding the probability of a rare event. From our point of view,
it seems much more reliable to assess that the probability of a feared event (failure of a system, natural disaster, etc.)
not exceeding a given level with high probability than to estimate the probability of this event
happening. 
Using Kriging metamodels to cope with the expensive black box model induces a random interpretation
of the probability to be estimated. Two strategies were studied in that context including our MBIS strategy. 
On a toy example, the efficiency of the two strategies was shown and it was highlighted that a sequential design sampling,
when possible, is preferable. 
Concerning the importance sampling strategy, further investigations could be about proposing an optimal splitting of the calls 
to the code used for the metamodel or for the importance sampling. Other concerns could be about tuning $\kappa$
to construct the set $\hat R_{\rho,\kappa}$.

We have dealt with a cross-validation method to assess the Kriging hypothesis. However, in the case
where the cross-validation leads one to reconsider this hypothesis, a solution is to extend the confidence
interval on the prediction by tuning by hand the parameter $\sigma^2$ in equation (\ref{krigeage}).
In Bayesian words, it can be called using a less informative prior distribution on $f$.

We have not managed to compute the posterior variance (\ref{postvariance})
by using a massive Monte Carlo integration in our examples since it is very small. However, other rare
event methods can be investigated since the variance no longer depends on $f$.

\bibliographystyle{apalike}
\bibliography{ABM11-RE}

\end{document}